\documentclass[12pt, a4paper]{article}
\usepackage[margin=1in]{geometry}
\usepackage{times}
\usepackage{amsmath, amssymb, graphicx, float}
\usepackage{algorithm, algpseudocode}
\usepackage[colorlinks=true]{hyperref}
\usepackage{array}
\usepackage{listings} 
\usepackage{xcolor} 

\usepackage{soul}

\begin{document}

\begin{center}
\textbf{\Large A Study Of Sudoku Solving Algorithms: \\Backtracking and Heuristic}

\vspace{12pt}

\normalsize Apekshya Bhattarai$^{1}$, Dinisha Uprety$^{1}$, Pooja Pathak$^{1}$, Safal Narshing Shrestha$^{1}$, Salina Narkarmi$^{1}$, Sanjog Sigdel$^{1*}$

\vspace{12pt}

\footnotesize $^{1}$Department of Computer Science, Kathmandu University, Dhulikhel, Nepal \\
$^{*}$Corresponding author: sanjog.sigdel@ku.edu.np
\end{center}

\begin{abstract}
\fontsize{10}{12}\selectfont
This paper presents a comparative analysis of Sudoku-solving strategies, focusing on recursive backtracking and a heuristic-based constraint propagation method. Using a dataset of 500 puzzles across five difficulty levels (Beginner to Expert), we evaluated performance based on average solving time. The heuristic approach consistently outperformed backtracking, achieving speedup ratios ranging from 1.27× in Beginner puzzles to 2.91× in Expert puzzles. These findings underscore the effectiveness of heuristic strategies, particularly in tackling complex puzzles across varying difficulty levels.
\end{abstract}


\section{Introduction}
Sudoku is a puzzle played on a partially filled 9x9 grid. The task is to complete
 the assignment using numbers from 1 to 9 such that the entries in each row, each
 column, and each major 3x3 block are pairwise different. Like for many logical
 puzzles, the challenge in Sudoku does not just lie in finding a solution [1]. Instead, it involves understanding the underlying strategies and techniques necessary for solving the puzzle efficiently. Players must navigate various constraints and make decisions based on limited information, often requiring critical thinking and pattern recognition. Moreover, the difficulty of Sudoku puzzles can vary significantly, demanding different levels of skill and problem-solving approaches. This multifaceted nature makes Sudoku not only an engaging activity but also a rich subject for algorithmic analysis and study.
 In the past few decades, distinct algorithms have been created for solving Sudoku puzzles. In this paper, we use Backtracking[2] and Heuristic[3] algorithms and compare their time to solve Sudoku puzzles.
 
    \hspace{3em} We have evaluated our methods on different sets of difficulties of the puzzles to see how efficient both algorithms were in comparison to each other and themselves, varying the difficulty. This evaluation gives a fair comparison of the difficulty of different puzzle sources. 
    
\subsection*{Backtracking}
This is the easiest approach as it avoids computational complexities. This algorithm visits all the empty cells in a specific order and fills it with a digit from 1-9, and checks its validity. If no choice is available for a cell then it backtracks and changes the digit of the previous cell. Thus the algorithm continues until all the cells are filled with appropriate digits. [2][4]

\subsection*{Heuristic}
This approach is an algorithm that combines human intuition and optimization to solve Sudoku puzzles [3]. It involves constraint propagation, where the algorithm eliminates impossible values for each cell based on existing constraints. It keeps track of possible candidates for each empty cell and uses techniques like naked singles or hidden pairs to progressively fill the grid. If no certain move is available, it may use backtracking with educated guesses, trying values and reverting if conflicts arise.

\section{Related Works}
The problem of solving Sudoku puzzles has been widely explored within the domain of Constraint Satisfaction Problems (CSPs) [5], particularly using backtracking combined with heuristics. These approaches aim to reduce the solution search space and improve the efficiency of solving time.

Simonis [1] proposed modeling Sudoku as a CSP using the all-different constraint, which ensures that each digit appears only once in every row, column, and 3×3 box. This approach laid the groundwork for constraint propagation techniques such as Arc Consistency (AC-3) and Forward Checking, which are widely used to prune inconsistent values early during the search. 

Norvig [6] introduced a well-known Python-based Sudoku solver that integrates backtracking with constraint propagation. This method uses the Minimum Remaining Value (MRV) heuristic to prioritize variables with the fewest legal values, significantly improving search efficiency. 

Rodrigues et al. improved heuristic performance by integrating both the Minimum Remaining Value (MRV) and Least Constraining Value (LCV) strategies [7].  Our solver adopts a similar strategy by implementing MRV, LCV, and other heuristics to optimize decision-making during solving.

Studies have experimented with hybrid techniques combining backtracking, constraint propagation, and domain-specific heuristics for enhanced performance [8]. Their work demonstrated the effectiveness of these methods in solving complex problems like Sudoku. Barker et al. [9] developed a method combining local search and constraint satisfaction, using backtracking along with a greedy approach to select promising variables, showing significant improvements for solving more challenging Sudoku puzzles.

In addition to CSP-based strategies, Sudoku solving has also been explored through the lens of combinatorial optimization. Sudoku is a special case of the Exact Cover problem, which is NP-Complete, as proven by the general problem of solving n×n Sudoku grids [10]. Knuth’s Dancing Links (DLX) algorithm [11], a technique for solving Exact Cover problems efficiently using sparse matrix manipulation, has been successfully applied to Sudoku solving. DLX performs particularly well due to its ability to efficiently undo and redo decisions during the backtracking process.

\section{Methodology}

We developed a web-based Sudoku application comprising two main sections: \textbf{Play} and \textbf{Solver}. The Play section features a custom puzzle generator that creates playable grids of varying difficulty. The Solver section allows users to input puzzles and compare two solving methods—recursive backtracking and heuristic-based constraint propagation—displaying their solving times side by side. This setup enabled real-time evaluation of algorithmic performance across multiple difficulty levels.

\subsection{Application Framework}
The architecture of our application adopts a modular design, with a clear separation between the frontend and backend components:
\begin{itemize}
    \item \textbf{Frontend}: Built using \texttt{HTML}, \texttt{CSS}, and \texttt{JavaScript}, offering a responsive and user-friendly interface.
    \item \textbf{Backend}: Developed using \texttt{Django} (Python), with \texttt{Postgres} as the database. Core logic and algorithm execution are handled server-side via standard Django APIs.
\end{itemize}

\subsection{Experimental Setup}
To evaluate the performance of both the recursive backtracking and heuristic-based Sudoku solving algorithms, we conducted experiments using a dataset of 500 Sudoku puzzles across various difficulty levels. These puzzles were generated through our own web-based Sudoku platform. The dataset was evenly categorized into five levels — Beginner, Easy, Medium, Hard, and Extreme — with 100 puzzles in each category. The difficulty classification was based on the number of initial clues (pre-filled cells) in each puzzle: \textbf{Beginner} puzzles contained 50 clues, \textbf{Easy} 40 clues, \textbf{Medium} 35 clues, \textbf{Hard} 27 clues, and \textbf{Extreme} 20 clues. The puzzles were collected using the solver section of our platform, and the solving time for both algorithms was recorded. \\
This experimental setup enabled a comprehensive analysis of algorithmic performance across varying levels of complexity, highlighting how each method scales with puzzle difficulty.

\subsection{Puzzle Generator}
To create a sudoku puzzle, we used backtracking method to completely fill an empty 9x9 grid. This process starts from the very first cell and proceeds cell by cell, ensuring that every placement complies with the rules of sudoku. Once a valid and fully populated grid was produced, we then applied the puzzle's difficulty by selectively removing numbers from the grid. The number of clues that remain corresponds to the chosen difficulty level; fewer clues generally mean a more challenging puzzle. By carefully "popping" values from the completed grid, we created a new playable puzzle, inviting solvers to engage with the logical challenge of filling in the missing numbers. The following pseudocode outlines the Sudoku puzzle generator that has been implemented: \newline 

\noindent \textbf{Input:} Empty 9×9 grid \\  
\textbf{Output:} Sudoku Puzzle based on the chosen difficulty

\begin{algorithm}[H]
\begin{algorithmic}[1]
\Function{IsValidMove}{$grid, row, col, num$}
    \If{$num$ is in $grid[row]$}
        \State \Return False
    \EndIf
    \If{$num$ is in column $col$ of $grid$}
        \State \Return False
    \EndIf
    \State $(start\_row, start\_col) \gets 3 \cdot (row \div 3), 3 \cdot (col \div 3)$
    \For{$i = 0$ to $2$}
        \For{$j = 0$ to $2$}
            \If{$grid[start\_row + i][start\_col + j] = num$}
                \State \Return False
            \EndIf
        \EndFor
    \EndFor
    \State \Return True
\EndFunction
\end{algorithmic}
\end{algorithm}
\begin{center}   \textbf{Fig.1 Pseudocode of ISVALIDMOVE function} \end{center}
\newpage

\begin{algorithm}
\begin{algorithmic}[1]
\Function{FindEmpty}{$grid$}
    \For{$i = 0$ to $8$}
        \For{$j = 0$ to $8$}
            \If{$grid[i][j] = 0$}
                \State \Return $(i, j)$
            \EndIf
        \EndFor
    \EndFor
    \State \Return None
\EndFunction
\end{algorithmic}
\end{algorithm}
\begin{center} \textbf{Fig.2 Pseudocode of FINDEMPTY function} \end{center}

\begin{algorithm}[H]
\begin{algorithmic}[1]
\Function{SolveSudoku}{$grid$}
    \State $(row, col) \gets$ \Call{FindEmpty}{$grid$}
    \If{no empty cell found}
        \State \Return True \Comment{Puzzle is solved}
    \EndIf
    \For{$num$ in shuffled $1$ to $9$}
        \If{\Call{IsValidMove}{$grid, row, col, num$}}
            \State $grid[row][col] \gets num$
            \If{\Call{SolveSudoku}{$grid$}}
                \State \Return True
            \EndIf
            \State $grid[row][col] \gets 0$ \Comment{Backtrack}
        \EndIf
    \EndFor
    \State \Return False
\EndFunction
\end{algorithmic}
\end{algorithm}
\begin{center} \textbf{Fig.3 Pseudocode of SOLVESUDOKU function} \end{center}

\begin{algorithm}[H]
\begin{algorithmic}[1]
\Function{GenerateSudoku}{$clues$}
    \State Initialize $grid$ as $9 \times 9$ zeros
    \For{$i$ in $\{0, 3, 6\}$}
        \State $nums \gets$ shuffled list from $1$ to $9$
        \For{$j = 0$ to $2$}
            \For{$k = 0$ to $2$}
                \State $grid[i + j][i + k] \gets$ \Call{Pop}{$nums$}
            \EndFor
        \EndFor
    \EndFor
    \State \Call{SolveSudoku}{$grid$}
    \State $solution \gets$ copy of $grid$
    \State $cells \gets$ all $(i, j)$ in grid
    \State Shuffle $cells$
    \For{first $81 - clues$ cells}
        \State $(row, col) \gets$ cell
        \State $grid[row][col] \gets 0$
    \EndFor
    \State \Return $(grid, solution)$
\EndFunction
\end{algorithmic}
\end{algorithm}
\begin{center} \textbf{Fig.4 Pseudocode of GENERATESUDOKU function} \end{center}

\subsection{Algorithm Design}
This research implemented two algorithms: heuristic and backtracking, to solve the puzzles generated by our generator. Below is an overview of each algorithm:

\subsubsection{Recursive Backtracking}
The backtracking algorithm systematically explores all possible configurations of the puzzle. It incrementally builds candidates for solutions and abandons a candidate as soon as it is determined that it cannot lead to a valid solution. \newline

\noindent
\textbf{Input:} 9×9 grid $G$ with some pre-filled values and zeros representing empty cells.\\  
\textbf{Output:} Solved grid or "No solution"

\begin{algorithm}[H]
\begin{algorithmic}[1]
\Function{Solve}{$G$}
    \For{$r \gets 0$ \textbf{to} $8$}
        \For{$c \gets 0$ \textbf{to} $8$}
            \If{$G[r][c] = 0$}
                \For{$num \gets 1$ \textbf{to} $9$}
                    \If{\Call{IsValid}{$G, r, c, num$}}
                        \State $G[r][c] \gets num$
                        \If{\Call{Solve}{$G$}} 
                            \State \Return \textbf{True}
                        \EndIf
                        \State $G[r][c] \gets 0$
                    \EndIf
                \EndFor
                \State \Return \textbf{False}
            \EndIf
        \EndFor
    \EndFor
    \State \Return \textbf{True}
\EndFunction
\end{algorithmic}
\end{algorithm}
\begin{center} \textbf{Fig.5 Pseudocode of SOLVE function} \end{center}

\textbf{Time Complexity :} \\
The results of calculations of the time complexity of the backtracking algorithm using the big theta notation is(average case)[12]  \[  
\boldsymbol{\theta}(n^3)  
\]

\subsubsection{Heuristic-Based Solver (MRV + LCV)}
The heuristic-based solver prioritizes selecting the variable with the fewest legal values remaining (MRV) and then chooses values that impose the least restriction on other variables (LCV), which significantly speeds up the solving process by reducing the search space and employs backtracking when a chosen value proves incorrect. \newline 

\noindent
\textbf{Input:} 9×9 grid $G$ with some pre-filled values and zeros representing empty cells.\\
\textbf{Output:} A completely filled valid Sudoku grid if solvable; otherwise, the message ``No solution''.

\newpage

\begin{algorithm}[H]
\begin{algorithmic}[1]
\Function{}{}{SolveSudokuWithHeuristics}{$(board)$}
    \State $results \gets$ \{solved: false, solution: null, error: null\}
    
    \State $possibilities \gets$ \Call{InitializePossibilities}{$board$}
    \State \Call{Heursitics}{$board, possibilities$}
    \State \Return $results$
\EndFunction
\end{algorithmic}
\end{algorithm}
\begin{center} \textbf{Fig.6 Pseudocode of SolveSudokuWithHeuristics function} \end{center}

\begin{algorithm}
\begin{algorithmic}[1]
\Function{Heuristics}{$board, possibilities$}

    \State $row, col, values \gets$ \Call{FindMostConstrainedCell}{$possibilities$}
    
    \If{$row = -1$ and $col = -1$}
        \State $results.solved \gets$ true
        \State $results.solution \gets$ copy of $board$
        \State \Return true
    \EndIf
    
    \If{$useRandomization$}
        \State Shuffle $values$
    \EndIf
    
    \State $cellCoord \gets (row, col)$
    
    \For{each $num$ in $values$}
        \If{\Call{IsValidMove}{$board, row, col, num$}}
            \State $board[row][col] \gets num$
            
            \State $updatedPossibilities \gets$ deep copy of $possibilities$
            \State Remove $cellCoord$ from $updatedPossibilities$
            
            \For{each $(r, c)$ in \Call{GetAffectedCells}{$row, col$}}
                \If{$(r, c)$ exists in $updatedPossibilities$}
                    \State Remove $num$ from $updatedPossibilities[(r, c)]$
                \EndIf
            \EndFor
            
            \If{\Call{Heuristics}{$board, updatedPossibilities$}}
                \State \Return true
            \EndIf
            
            \State $board[row][col] \gets 0$  \Comment{Backtrack}
        \EndIf
    \EndFor
    
    \State $possibilities[cellCoord] \gets$ set of $values$
    \State \Return false
\EndFunction

\end{algorithmic}
\end{algorithm}
\begin{center} \textbf{Fig.7 Pseudocode of Heuristics function} \end{center}

\newpage

\begin{algorithm}[H]
\begin{algorithmic}[1]
\Function{GetAffectedCells}{$(row, col)$}
\State Initialize empty set \texttt{affected}
\For{$c \gets 0$ to $8$}
    \If{$c \neq col$}
        \State Add $(row, c)$ to \texttt{affected}
    \EndIf
\EndFor
\For{$r \gets 0$ to $8$}
    \If{$r \neq row$}
        \State Add $(r, col)$ to \texttt{affected}
    \EndIf
\EndFor
\State $box\_row \gets 3 \times \lfloor row / 3 \rfloor$
\State $box\_col \gets 3 \times \lfloor col / 3 \rfloor$
\For{$r \gets box\_row$ to $box\_row + 2$}
    \For{$c \gets box\_col$ to $box\_col + 2$}
        \If{$(r, c) \neq (row, col)$}
            \State Add $(r, c)$ to \texttt{affected}
        \EndIf
    \EndFor
\EndFor
\State \Return \texttt{affected}
\EndFunction
\end{algorithmic}
\end{algorithm}
\begin{center} \textbf{Fig.8 Pseudocode of GETAFFECTEDCELLS function} \end{center}

\begin{algorithm}
\begin{algorithmic}[1]
\Function{IsValidMove}{$board, row, col, num$}
    \For{$x \gets 0$ to $8$}
        \If{$board[row][x] = num$ or $board[x][col] = num$}
            \State \Return False
        \EndIf
    \EndFor
    \State Compute box start indices
    \For{each cell in the 3x3 box}
        \If{cell has value $= num$}
            \State \Return False
        \EndIf
    \EndFor
    \State \Return True
\EndFunction
\end{algorithmic}
\end{algorithm}
\begin{center} \textbf{Fig.9 Pseudocode of IsValidMove function} \end{center}
\newpage

\begin{algorithm}
\begin{algorithmic}[1]
\Function{InitializePossibilities}{$(board)$}
\State Initialize empty dictionary \texttt{possibilities}
\For{each $(row, col)$ in board}
    \If{$board[row][col] = 0$}
        \State \texttt{possibilities}[$(row, col)$] $\gets \{1,\dots,9\}$
    \EndIf
\EndFor
\For{each $(row, col)$ in board}
    \If{$board[row][col] \neq 0$}
        \State $value \gets board[row][col]$
        \For{each $(r, c)$ in GetAffectedCells(row, col)}
            \If{$(r, c) \in$ \texttt{possibilities}}
                \State Remove $value$ from \texttt{possibilities}[$(r, c)$]
            \EndIf
        \EndFor
    \EndIf
\EndFor
\State \Return \texttt{possibilities}
\EndFunction
\end{algorithmic}
\end{algorithm}
\begin{center} \textbf{Fig.10 Pseudocode of INITIALIZEPOSSIBILITIES function} \end{center}

\begin{algorithm}
\begin{algorithmic}[1]
\Function{FindMostConstrainedCell}{$(possibilities)$}
\If{\texttt{possibilities} is empty}
    \State \Return $(-1, -1, [])$
\EndIf
\State $min \gets 10$, $best\_cell \gets (-1, -1)$
\For{each $(row, col), values$ in \texttt{possibilities}}
    \If{$|values| < min$}
        \State $min \gets |values|$
        \State $best\_cell \gets (row, col)$
        \State $best\_values \gets values$
    \EndIf
\EndFor
\State \Return $(row, col, list(best\_values))$\
\EndFunction
\end{algorithmic}
\end{algorithm}
\begin{center} \textbf{Fig.11 Pseudocode of FINDMOSTCONSTRAINEDCELL function} \end{center}

\newpage

\textbf{Time Complexity :} \\
The time complexity of the heuristic algorithm for solving a Sudoku puzzle is almost in polynomial time. [13] \\

\textbf{Heuristic Cost Function :}\\
A heuristic cost function is a mathematical function used to estimate the cost or effort required to reach a solution from a given state in a problem space. 

\[  
f(s) = g(s) + h(s)  
\]  

\begin{itemize} 
\item $ f(s) $ : Total estimated cost 
\item  $ g(s) $ = Cost value of the trajectory performed from the root to the current ($s$) state;
\item $ h(s) $ = Heuristic function that estimates the cost of the
path from the current state to the destination state. [14]
\end{itemize}  

\noindent In our approach, these are defined as:\\
A heuristic cost function \( f(s) \) to evaluate the quality of a partially completed Sudoku state \( s \). This function guides the backtracking algorithm by combining both the progress made and the difficulty of remaining decisions.

\begin{itemize}
    \item \( g(s) = 81 - |\text{possibilities}| \) \\
    where \( |\text{possibilities}| \) is the number of currently empty cells.

    \item \( h(s) = \sum_{(i,j) \in \text{possibilities}} \frac{1}{|P(i,j)|} \) \\
    where \( P(i,j) \) is the set of possible values for cell \( (i,j) \).
\end{itemize}

\noindent Therefore, the full heuristic becomes:

\[
f(s) = 81 - |\text{possibilities}| + \sum_{(i,j)} \frac{1}{|P(i,j)|}
\]

\section{Result}

To evaluate the effectiveness of the two solving approaches, we ran both algorithms on 500 Sudoku puzzles spanning five difficulty levels: Beginner, Easy, Medium, Hard, and Expert. The table below summarizes the average solving time (in milliseconds) for each method.

\begin{table}[H]
    \centering
    \textbf{\caption{Average Solving Time for 500 Sudoku Puzzles}}
    \begin{tabular}{|l|c|c|}
        \hline
        \textbf{Difficulty} & \textbf{Backtracking (ms)} & \textbf{Heuristic (ms)} \\
        \hline
        Beginner & 0.93 & 0.73 \\
        Easy     & 4.56 & 3.27 \\
        Medium   & 14.78 & 8.26 \\
        Hard     & 173.45 & 78.88 \\
        Expert   & 360.72 & 123.80 \\
        \hline
    \end{tabular}
\end{table}

\subsection{Speedup Analysis}

To better understand the relative improvement of the heuristic solver, we computed the speedup ratio[15] as the time taken by the backtracking approach divided by the time taken by the heuristic approach. The heuristic solver consistently outperforms backtracking, with its advantage becoming more significant at higher difficulty levels.

\begin{table}[H]
\centering
\textbf{\caption{Speedup Ratio of Heuristic Solver Over Backtracking}}
\begin{tabular}{|l|c|c|c|l|}
\hline
\textbf{Difficulty} & \textbf{Backtracking (ms)} & \textbf{Heuristic (ms)} & \textbf{Ratio (y/x)} & \textbf{Interpretation} \\
\hline
Beginner & 0.93 & 0.73 & 1.27 & $\sim$1.27x faster \\
Easy     & 4.56 & 3.27 & 1.39 & $\sim$1.39x faster \\
Medium   & 14.78 & 8.26 & 1.79 & $\sim$1.79x faster \\
Hard     & 173.45 & 78.88 & 2.20 & $\sim$2.20x faster \\
Expert   & 360.72 & 123.80 & 2.91 & $\sim$2.91x faster \\
\hline
\end{tabular}
\end{table}

\begin{figure}[H]
    \centering
    \includegraphics[width=0.85\linewidth]{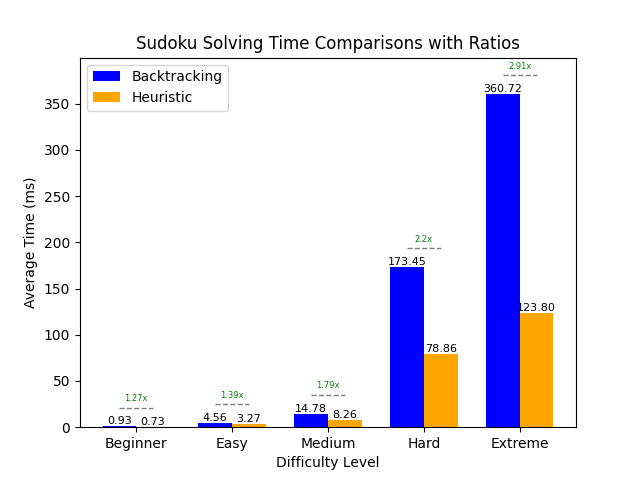}
\end{figure}
\begin{center} \textbf{Fig.12 Barchart Solving Time Comparison Across Difficulties } \end{center}

\begin{figure}[H]
    \centering
    \includegraphics[width=0.85\linewidth]{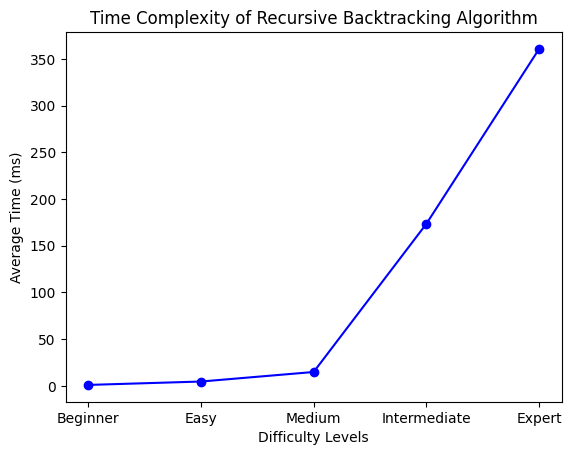}
\end{figure}
\begin{center} \textbf{Fig.13 Line Chart Solving Time of Backtracking Across Difficulties} \end{center}

\begin{figure}[H]
    \centering
    \includegraphics[width=0.85\linewidth]{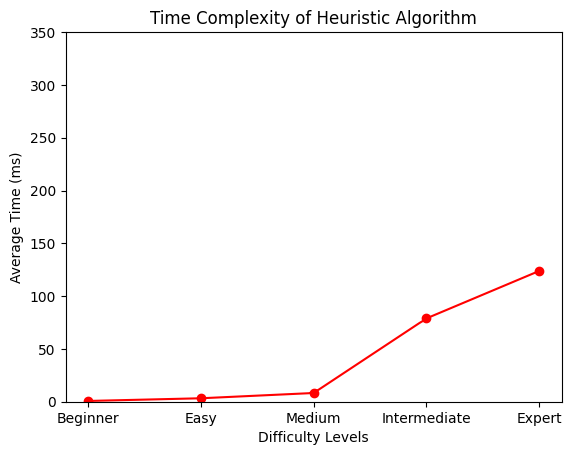}
\end{figure}
\begin{center} \textbf{Fig.13 Line Chart Solving Time of Heuristic Across Difficulties} \end{center}

\begin{figure}[H]
    \centering
    \includegraphics[width=0.85\linewidth]{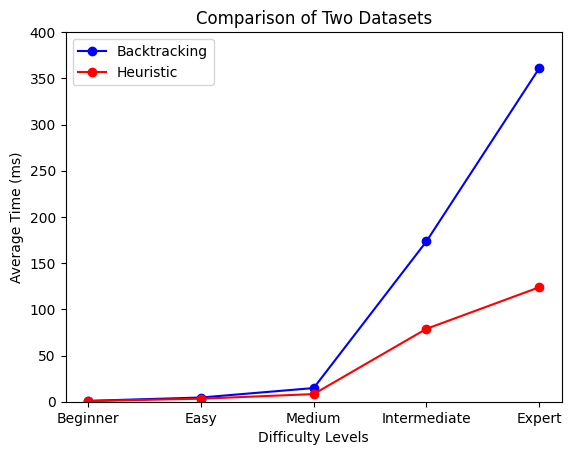}
\end{figure}
\begin{center} \textbf{Fig.13 Line Chart Solving Time Comparison Across Difficulties} \end{center}

The analysis shows that as puzzle difficulty increases, the gap in performance between the two methods widens. The \textbf{speedup ratios} reveal that the heuristic method becomes more advantageous for harder puzzles. Additionally, the heuristic solver maintains \textbf{consistency}, performing efficiently across all difficulty levels. Overall, while backtracking serves as a simple baseline, the heuristic method significantly improves performance, particularly with more complex puzzles.

\section{Conclusion}
Heuristic-based methods greatly improved the effectiveness of Sudoku solvers, especially when dealing with more challenging puzzles. Although recursive backtracking is a straightforward and efficient technique, the heuristic approach consistently surpasses it, demonstrating significant speed advantages at every difficulty level. As noted in section 2, there are additional methods that can provide various performance attributes for solving a Sudoku puzzle. Our results emphasize the scalability and effectiveness of heuristic methods, particularly in tackling more complex puzzles. This study provides important understanding of the relative strengths of two different techniques and indicates that heuristic methods are particularly effective compared to backtracking for efficiently solving challenging Sudoku puzzles.

\section*{References}
\begin{enumerate}
    \item Simonis, H. (2005, October). Sudoku as a constraint problem. \textit{In CP Workshop on modeling and reformulating Constraint Satisfaction Problems} (Vol. 12, pp. 13-27). Sitges, Spain: Citeseer.
    \item Christopher, J. (2017). Solving Sudoku Puzzles using Backtracking Algorithms.
    \item Pillay, Nelishia. (2012). Finding Solutions to Sudoku Puzzles Using Human Intuitive Heuristics. \textit{SACJ}. 49. 25-34. 10.18489/sacj.v49i0.111. 
    \item Chatterjee, S., Paladhi, S., \& Chakraborty, R. A Comparative Study On The Performance Characteristics Of Sudoku Solving Algorithms. \textit{IOSR Journals (IOSR Journal of Computer Engineering)}, 1(16), 69-77.\
    \item B. Crawford, M. Aranda, C. Castro and E. Monfroy, "Using Constraint Programming to solve Sudoku Puzzles," \textit{2008 Third International Conference on Convergence and Hybrid Information Technology}, Busan, Korea (South), 2008, pp. 926-931, doi: 10.1109/ICCIT.2008.154.
    \item Norvig, P. (2006). Solving Every Sudoku Puzzle.  \textit{https://norvig.com/sudoku.html}
    \item Rodrigues, C., Galvão, E., \& Azevedo, R. LSVF: A heuristic search to reduce the backtracking calls when solving Constraint Satisfaction Problems. SI nforme, 19.
    \item Kumar, V. (1992). Algorithms for Constraint-Satisfaction Problems: A Survey. \textit{AI Magazine}, 13(1), 32–44.
    \item C. Barker, J. Bridge, and P. Nightingale, “Combining Heuristics, Local Search and Backtracking for Sudoku,” in UK Workshop on Computational Intelligence, 2013.
    \item Yato, T., \& Seta, T. (2003). Complexity and completeness of finding another solution amand its application to puzzles. \textit{IEICE Transactions on Fundamentals of Electronics, Communications and Computer Sciences}, E86-A(5), 1052–1060.
    \item Knuth, D.E. (2000). Dancing Links. \textit{Journal of Algorithms}, 45(2), 123-145.
    \item Sitorus, P., \& Zamzami, E. M. (2020, June). An implementation of backtracking algorithm for solving a Sudoku-Puzzle based on Android. In Journal of Physics: Conference Series (Vol. 1566, No. 1, p. 012038). IOP Publishing.
    \item Chen, Z. (2009). Heuristic reasoning on graph and game complexity of sudoku. \textit{arXiv preprint arXiv:0903.1659}.
    \item Silva, J. B. B., Siebra, C. A., \& Nascimento, T. P. (2016). A simplified cost function heuristic applied to the A*-based path planning. \textit{International Journal of Computer Applications in Technology}, 54(2), 96. 
    \item Pearl, J. (1984). \textit{Heuristics: Intelligent Search Strategies for Computer Problem Solving} (pp. 112–115). Addison-Wesley.
\end{enumerate}

\end{document}